\documentclass[aps,prl,twocolumn,superscriptaddress]{revtex4-2}
\usepackage{graphicx,color}

\begin{document}


\title{Fermi arcs dominating the electronic surface properties of trigonal PtBi$_2$}



\author{Sven Hoffmann}
\affiliation{Bergische Univertsität Wuppertal, 42119 Wuppertal, Germany }
\affiliation{Leibniz Institute for Solid State and Materials Research, Helmholtzstra{\ss}e 20, 01069 Dresden, Germany}

\author{Sebastian Schimmel}
\affiliation{Bergische Univertsität Wuppertal, 42119 Wuppertal, Germany }
\affiliation{Leibniz Institute for Solid State and Materials Research, Helmholtzstra{\ss}e 20, 01069 Dresden, Germany}

\author{Riccardo Vocaturo}
\affiliation{Leibniz Institute for Solid State and Materials Research, Helmholtzstra{\ss}e 20, 01069 Dresden, Germany}

\author{Joaquin Puig}
\affiliation{Leibniz Institute for Solid State and Materials Research, Helmholtzstra{\ss}e 20, 01069 Dresden, Germany}
\affiliation{Centro Atómico Bariloche and Instituto Balseiro, CNEA,CONICET and Institutode Nanociencia y Nanotecnología , 8400 San Carlos de Bariloche, Argentina}

\author{Grigory Shipunov}
\affiliation{Leibniz Institute for Solid State and Materials Research, Helmholtzstra{\ss}e 20, 01069 Dresden, Germany}

\author{Oleg Janson}
\affiliation{Leibniz Institute for Solid State and Materials Research, Helmholtzstra{\ss}e 20, 01069 Dresden, Germany}

\author{Saicharan Aswartham}
\affiliation{Leibniz Institute for Solid State and Materials Research, Helmholtzstra{\ss}e 20, 01069 Dresden, Germany}

\author{Danny Baumann}
\affiliation{Leibniz Institute for Solid State and Materials Research, Helmholtzstra{\ss}e 20, 01069 Dresden, Germany}

\author{Bernd B\"uchner}
\affiliation{Leibniz Institute for Solid State and Materials Research, Helmholtzstra{\ss}e 20, 01069 Dresden, Germany}
\affiliation{Institute of Solid State Physics, Technische Universit{\"a}t Dresden, 01069 Dresden,Germany}
\affiliation{Center for Transport and Devices, Technische Universit{\"a}t Dresden, 01069 Dresden, Germany}

\author{Jeroen van den Brink}
\affiliation{Leibniz Institute for Solid State and Materials Research, Helmholtzstra{\ss}e 20, 01069 Dresden, Germany}

\author{Y. Fasano}
\affiliation{Leibniz Institute for Solid State and Materials Research, Helmholtzstra{\ss}e 20, 01069 Dresden, Germany}
\affiliation{Centro Atómico Bariloche and Instituto Balseiro, CNEA,CONICET and Institutode Nanociencia y Nanotecnología , 8400 San Carlos de Bariloche, Argentina}

\author{Jorge I. Facio}
\affiliation{Centro Atómico Bariloche and Instituto Balseiro, CNEA,CONICET and Institutode Nanociencia y Nanotecnología , 8400 San Carlos de Bariloche, Argentina}

\author{C. Hess}
\email[]{c.hess@uni-wuppertal.de}
\affiliation{Bergische Univertsität Wuppertal, 42119 Wuppertal, Germany }


\date{\today}

\begin{abstract}

Materials combining topologically non-trivial behavior and superconductivity offer a potential route for quantum computation. However, the set of available materials intrinsically realizing these properties are scarce. Recently, surface superconductivity has been reported in PtBi$_2$ in its trigonal phase and an inherent Weyl semimetal phase has been predicted. Here, based on scanning tunneling microscopy experiments, we reveal the signature of topological Fermi arcs in the normal state patterns of the quasiparticle interference. We show that the scattering between Fermi arcs dominates the interference spectra, providing conclusive evidence for the relevance of Weyl fermiology for the surface electronic properties of trigonal PtBi$_2$.

\end{abstract}


\maketitle

\section{Introduction}

Recently, the trigonal phase of PtBi$_2$ (t-PtBi$_2$) has been established as a significant candidate for studying the emergence of superconductivity in a topological Weyl semimetal. In addition to hosting Weyl nodes near the Fermi energy according to ab-initio calculations \cite{Veyrat2023}, transport measurements in t-PtBi$_2$ revealed superconductivity at very low temperatures ($T\lesssim1$~K) \cite{Shipunov2020,Zabala2024} including signatures of a Berezinskii-Kosterlitz-Thouless transition \cite{Berezinskii1971,Kosterlitz1973} for samples with thicknesses of tens of nanometers \cite{Veyrat2023}. Furthermore, surface superconductivity at even higher temperatures of 5~K was revealed by scanning tunneling spectroscopy (STS) at the surface of t-PtBi$_2$ samples \cite{Schimmel2023}, and angle resolved photoemission spectroscopy (ARPES) located  superconductivity at the topological Fermi arc states of this material \cite{Kuibarov2024}.
This feature suggests that exclusively the topological surface states, rather than the bulk states, become superconducting, which would render t-PtBi$_2$ truly exceptional among other superconducting Weyl semimetals \cite{Bachmann2017,Qi2016,Xing2020,Wang2021}.
Thus, in view of the multiband nature of t-PtBi$_2$, established by  ARPES \cite{Kuibarov2024,Thirupathaiah2018} as well as by quantum oscillation \cite{Gao2018} studies, a crucial question is the relevance of all those bulk states for the surface electronic properties.


In order to address this issue, we investigate in this work the quasiparticle interference (QPI) \cite{Hoffman2002,Zheng2018} of normal state t-PtBi$_2$ by means of scanning tunneling microscopy and spectroscopy (STM/STS) experiments and calculations of the joint density of states (JDOS) based on ab-initio density-functional theory (DFT). We identify characteristic structures which dominate the Fourier transformed QPI with excellent agreement between experiment and theory, and we show that these structures arise from scattering processes between the topological Fermi arcs. Thus, our joint experimental and theoretical approach confirms the Weyl semimetallic nature of t-PtBi$_2$ and clarifies that the surface electronic properties indeed are governed by the Weyl surface states.

\section{Results}
      
\begin{figure*}
\includegraphics[scale=0.4]{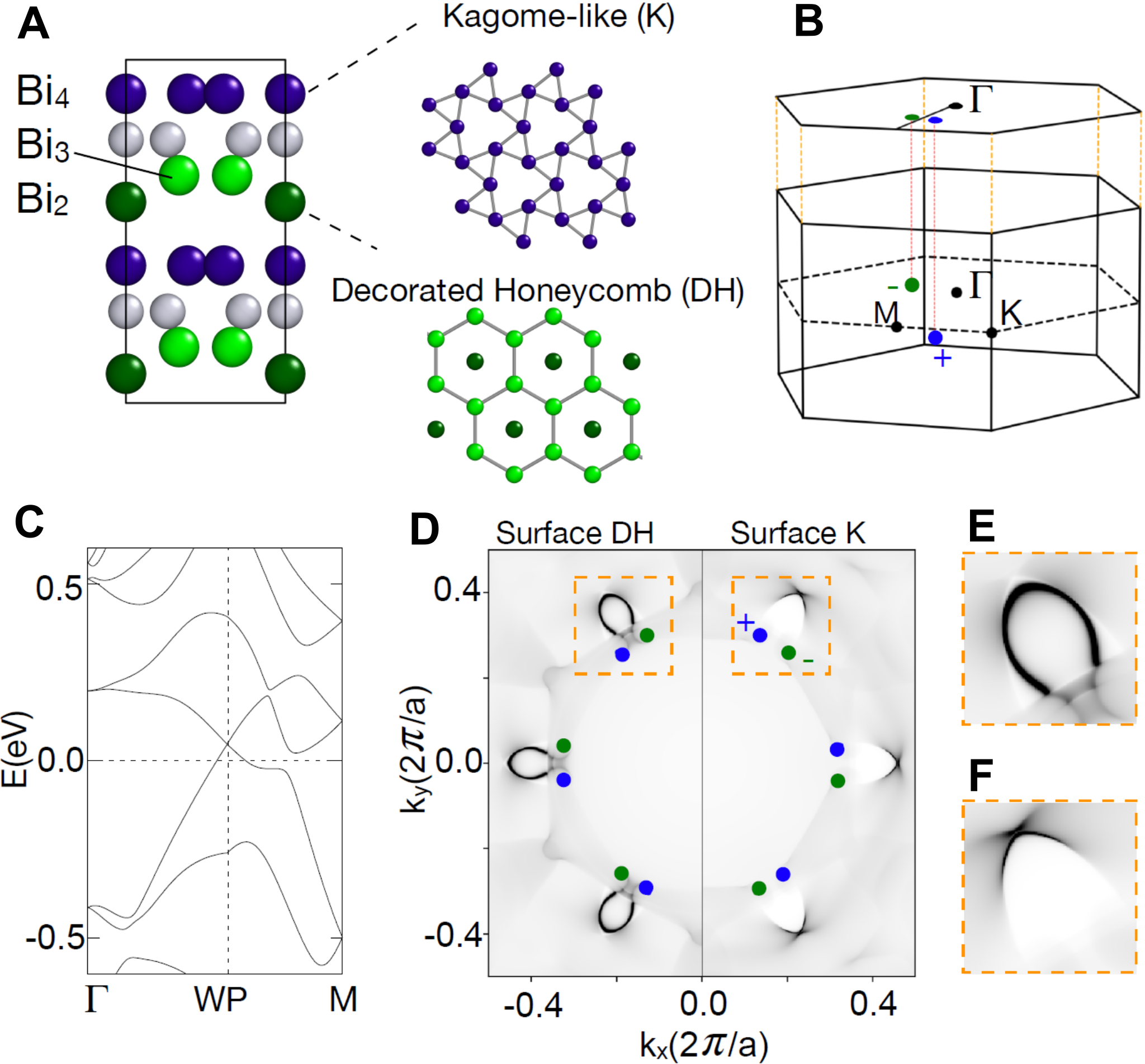}
\caption{(A) Crystal structure, with a detail of the planes formed by Bi at different Wyckoff positions. Bi$_4$ atoms form a kagome-like lattice while Bi$_2$ and Bi$_3$ a decorated honeycomb structure. (B) Brillouin zone and position of a Weyl node pair. (C) Bulk electronic structure. WP is the Weyl point at (0.324,  $\pm$ 0.041, $\pm$ 0.153) (in units of 2$\pi$/a) (D) Surface spectral density at the Fermi energy. Left- and right- halves of the plot correspond to the DH and K terminations. (E,F) Zooms into the Fermi arcs for both cases. }
\label{fig:Figure1}
\end{figure*}

The material t-PtBi$_2$ (with crystal symmetry SG 157) consists of a stack of Pt and Bi-layers where each atomic plane has three-fold rotation and reflection symmetries. The Bi-layers above and below the Pt planes have different arrangements of Bi atoms, effectively breaking the inversion symmetry. If we consider the shortest Bi-Bi separations within each Bi-layer as bonds, one of them features a kagome-like structure (K), while the other is a decorated honeycomb lattice (DH), see Fig.~1A. The respective termination has a strong influence on the surface electronic structure, that we describe here based on relativistic DFT within the generalized gradient approximation (see "Methods"). Previous calculations of the bulk electronic structure predict the existence of Weyl nodes located approximately 50 meV above the Fermi energy connecting the highest-lying valence band with the lowest-lying polarization band \cite{Veyrat2023}. Point symmetries together with the time-reversal symmetry lead to 12 such crossings, see a single pair in Fig.~1B. The bulk energy dispersion near the topological band crossings corresponds to type-I Weyl nodes (Fig.~1C).  A priori the relevance of Weyl nodes to the electronic structure at the boundaries in a semimetal where several bands form bulk Fermi surfaces is not obvious. To address this point, we performed calculations for semi-infinite slabs with the surface perpendicular to the [001] direction, both for DH and K surfaces. As shown in Fig.~1D-F, Fermi arcs can be seen at the Fermi energy.  Note that the momentum dependence of spectral weight is sensitive to the surface termination. For the DH surface, elongated Fermi arcs are discernible, see Fig.~1E. This contrasts with the case of the K surface where only the apex is observed, see Fig.~1F. According to our calculations, type K Fermi arcs maintain their shape over a large energy range and substantial changes become apparent only at energies differing more than about 25 meV from the Fermi energy. Fermi arcs at the DH termination show a similar behavior above the Fermi level whereas they acquire a notable dispersion below it. (Supplementary Fig.~S1 and S2).

\begin{figure*}
\includegraphics[scale=0.7]{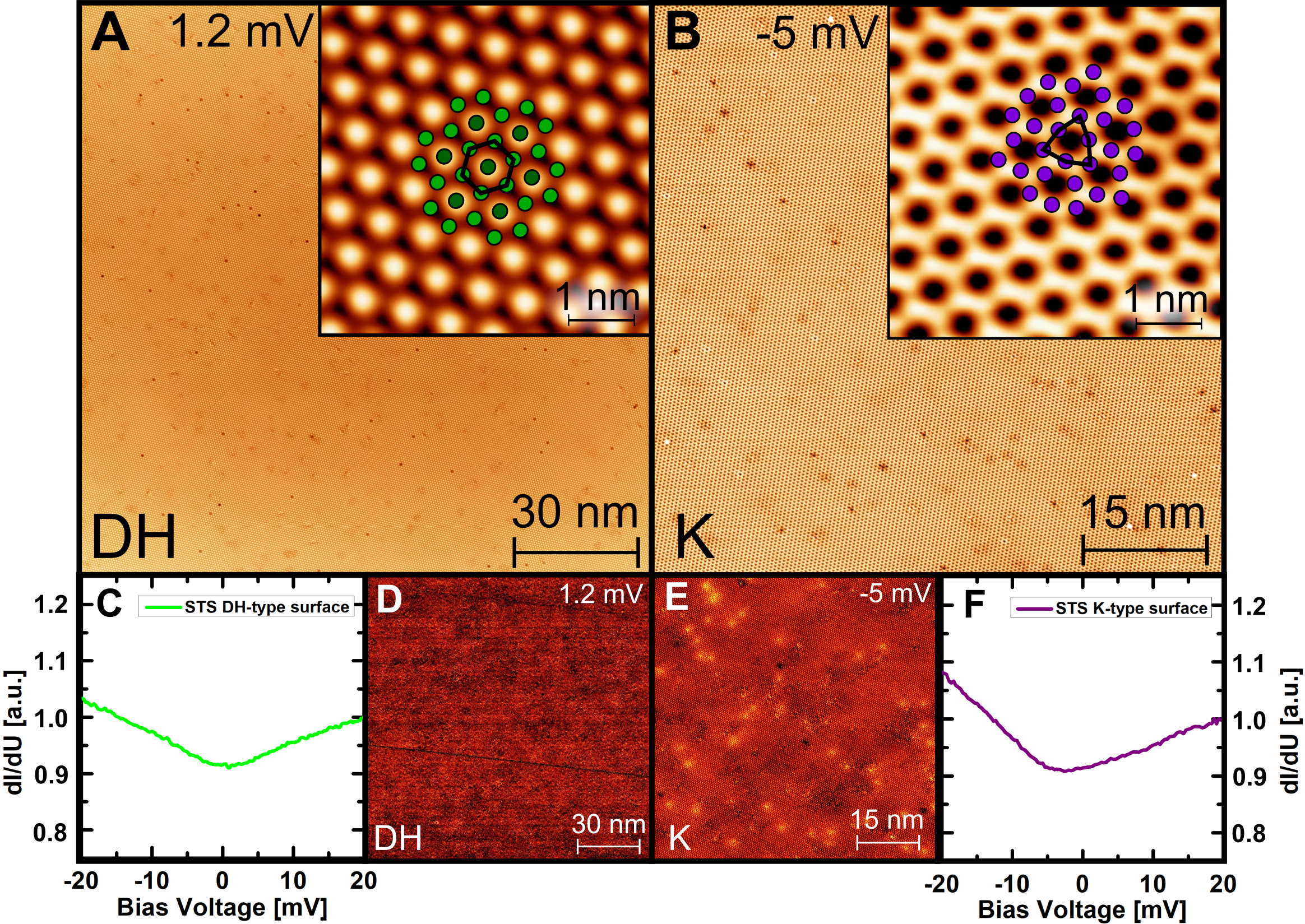}
\caption{(A) STM topography (150x150 nm, $U=1.2$ mV, $I= 100$  pA, $U_{\text{rms}}=$ 0.5 meV, $T=5$ K) showing DH surface (sample 2). Dark and light green circles represent the positions of Bi$_2$ and Bi$_3$ atoms, respectively.   (B) STM topography (75x75 nm, $U= -5$ mV, $I= 500$  pA, $U_{\text{rms}}= 3$ meV, $T= 30$ mK) showing a K surface (sample 1). Purple circles represent the position of Bi4 atom. Insets in A and B are 5x5 nm topographies depicting the commonly observed atomic corrugation. (D) and (E) show the differential conductance maps recorded simultaneously with A and B. Single point differential conductance spectra recorded on the sample surfaces shown in panel A and B are presented in (C) ($T=5$ K) and (F) ($T=30$ mK), respectively. They do not exhibit signs of a gap around the Fermi level indicating the absence of a superconducting transition.  }
\label{fig:Figure2}
\end{figure*}

Experimentally, both terminations can be realized by cleaving the samples between van der Waals bound Bi-layers \cite{Schimmel2023,Nie2020}. Fig.~2A, B shows typical STM atomic-resolution topographies in DH and K terminations. In topographies of DH surfaces primarily the protruding Bi$_2$ atoms forming a hexagonal structure are imaged as indicated in Fig.~2A. In contrast, in the K termination the negative of a hexagonal structure is observed, associated with the characteristic voids of the kagome-like lattice, see schematics in inset of Fig.~2B.

Previous work revealed that the signatures of surface superconductivity in tunneling experiments are subtle since the measured gap magnitude often is spatially inhomogeneous and sample dependent. This includes the complete absence of the surface superconductivity on certain sample areas \cite{Schimmel2023}.
Indeed, STS on the surfaces shown in Fig.~2A and B, measured at 5~K and 30 mK, respectively, shows no sign of superconductivity (see Fig.~2C, F).
Thus, these surfaces are ideal for studying the normal state Fermiology of t-PtBi$_2$. To this end, we scrutinize high-resolution maps of the differential conductance ($dI/dU$) which were simultaneously registered together with the shown topography data, see panels D, E for the real space data and Fig.~3A and B for the corresponding Fourier transforms. The bias voltages for the maps (1.2~meV and -5~meV) are close to zero bias, i.e., the Fermi level.
Furthermore, at low temperatures, $dI/dU$ is proportional to the local density of state (LDOS) induced by impurities at the surface; therefore, we expect the data to allow comparison with the calculated spectral density shown in Fig. 1. To this end, a particularly useful approach is to compute (and compare to) the joint density of states (JDOS), which is related to the surface spectral function $A(\omega, \mathbf k)$ by:

\begin{equation}
JDOS(\omega,\mathbf q)=\int{d^2\mathbf k A(\omega,\mathbf k) A(\omega,\mathbf k-\mathbf q)}.
\label{eq:JDOS}
\end{equation}


The reciprocal space QPI patterns in Fig.~3A and B exhibit a very rich structure which is very distinct for the two surface terminations. For the DH surface (Fig.~3A), the most prominent structures apart the Bragg peaks stemming from the atomic corrugation (indicated by $\mathbf q_a$) are a clover leaf-like pattern (red triangle) and a doublet (orange ellipse) located in the $\Gamma-K$ and the $\Gamma-M$ direction, respectively. The doublet is also present for the K surface in Fig.~3B (orange ellipse), but instead of the clover leaf, there is a triplet of high intensity spots, and there is another triplet further out along the $\Gamma-K$ direction (red and pink triangles, respectively). Since the QPI is directly related with the electron scattering in the material, we assign scattering vectors $\mathbf q_1^{DH}$ to $\mathbf q_5^{DH}$  and $\mathbf q_1^K$ to $\mathbf q_8^K$ to the highest conductance locations within these structures on the DH and K terminations.

\begin{figure*}
\includegraphics[scale=0.7]{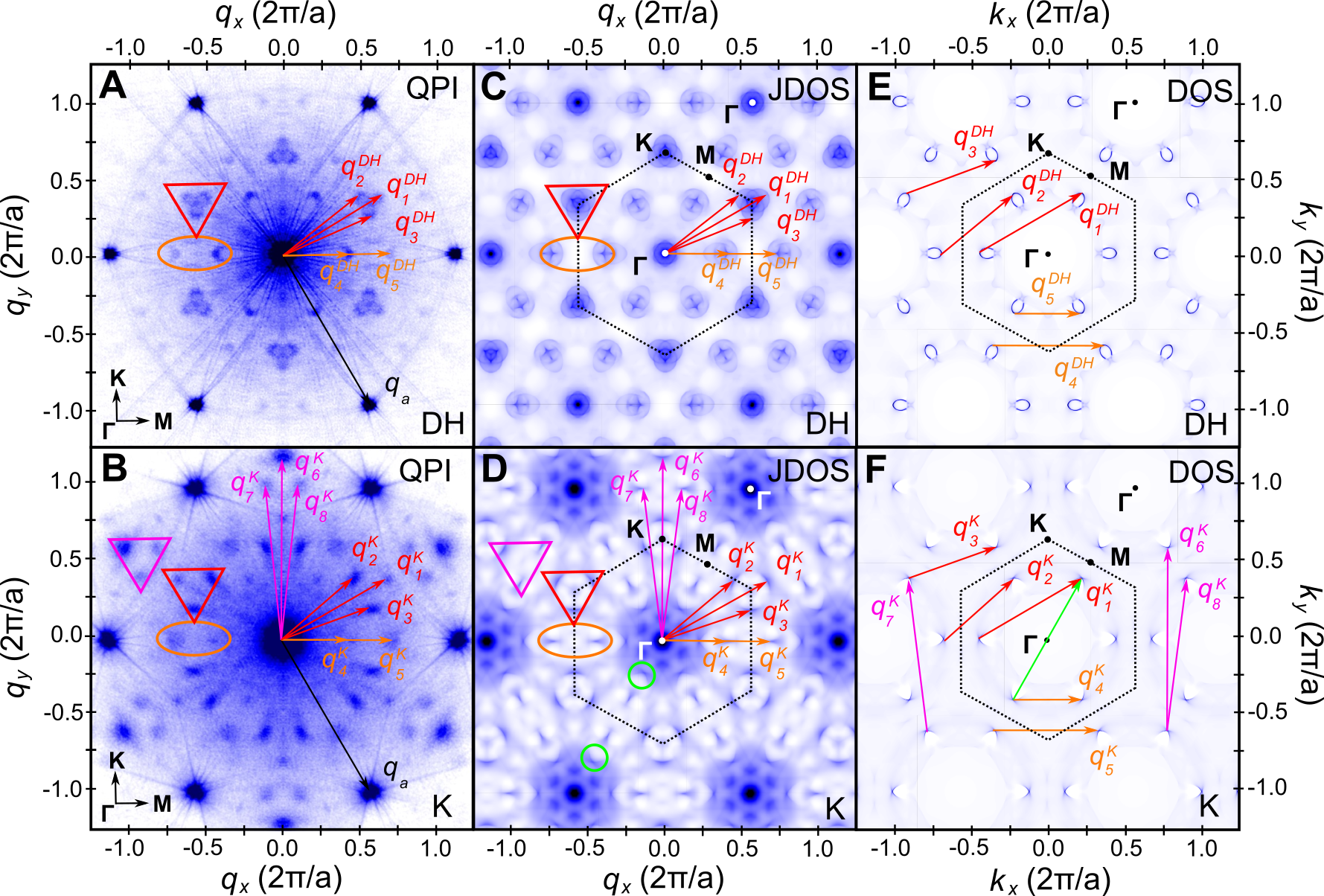}
\caption{(A,B) QPI for a DH surface at 1.2 meV and type K at -5 meV derived though Fourier transformation from differential conductance maps shown in Fig.~2 D and E.  Scattering vectors $\mathbf q_i$, marked by colored arrows indicate interference peaks in the QPI data.  The red and pink triangles and the orange ellipse mark features associated with scattering between Fermi arcs.  (C,D)  JDOS for DH and K terminations at 0 meV and -5 meV, respectively. Scattering vectors identical to those in A and B have been superimposed onto the JDOS. Green circles mark scattering vectors connecting time-reversal pairs of Fermi arcs. The outer circle corresponds to the green arrow in panel F, while the inner circle is related the outer one by a translation vector plus time-reversal symmetry.  (E,F) DFT-based constant energy contour of DH and K surfaces at the respective measurements energy. The white dashed line marks the first Brillouin zone. The QPI data has been symmetrized according to point and time-reversal symmetries.  }
\label{fig:Figure3}
\end{figure*} 

The experimental QPI pattern shows a striking resemblance with the JDOS computed from our DFT calculations, as evident in the comparison between data in  Fig.~3A, C and B, D. This good agreement in both surface terminations is also found in a wide energy range as discussed later.
%
%
Based on the good agreement obtained, we now try to infer which electronic states are involved in the dominant scattering processes by comparing the associated wave vectors with the structures in the spectral density at the Fermi energy (labeled ``DOS'' in Fig.~3E, F). Fig.~3E includes the wave vectors  $\mathbf q^{DH}_1$ to $\mathbf q^{DH}_5$ and it can be seen that these perfectly connect Fermi arcs. Similar results are found for the K surface, see $\mathbf q^K_1$ to $\mathbf q^K_8$  in Fig.~3F. Already these coarse comparisons provide strong evidence that the QPI of t-PtBi$_2$ is governed by electron scattering between the topological Fermi arcs.

\begin{figure*}
\includegraphics[scale=0.7]{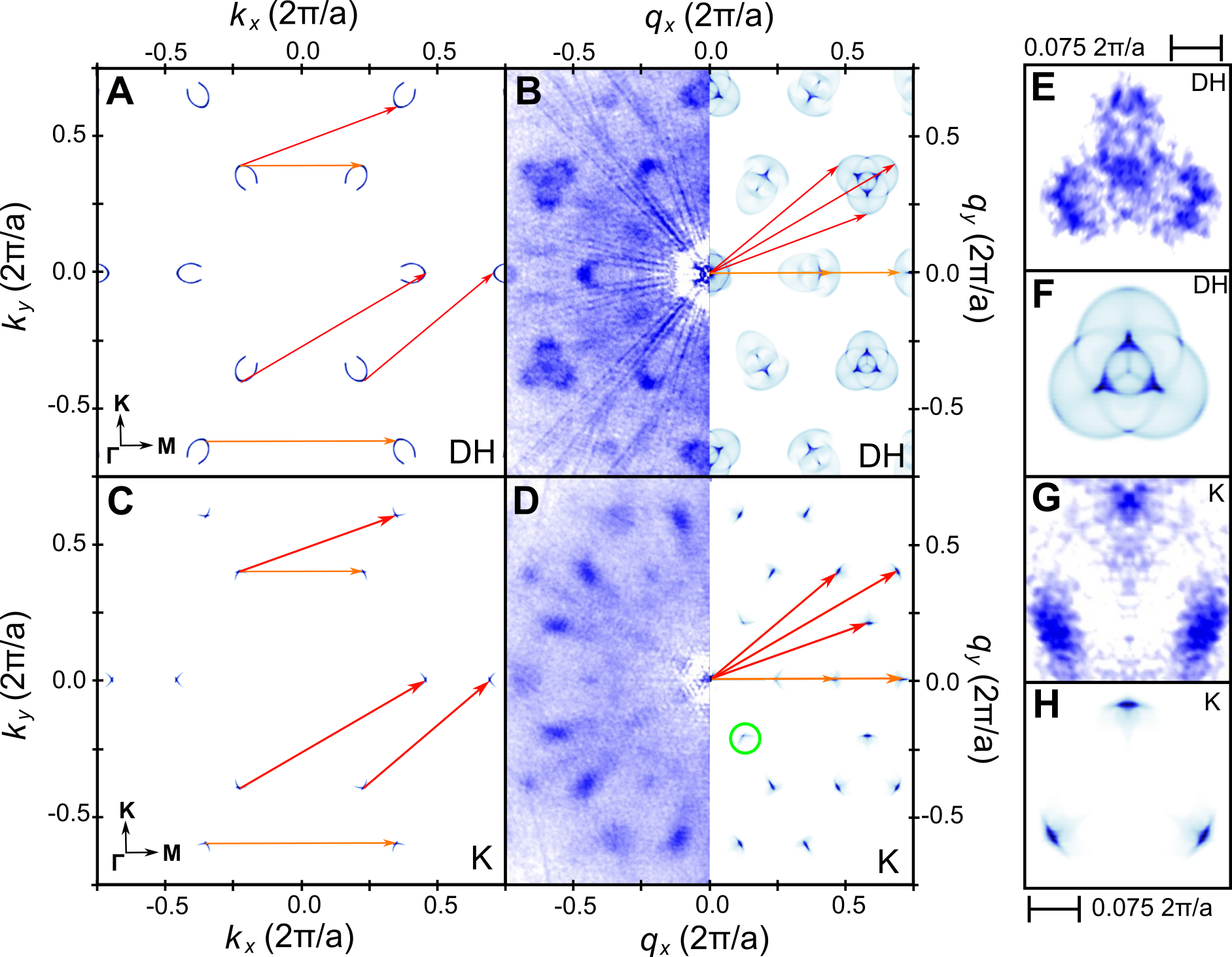}
\caption{(A,B) Simplified spectral density consisting only of the Fermi arcs for DH and K surfaces. (C,D) QPI and JDOS for DH and K surfaces. The JDOS is based on the data in panels A and B. Arrows indicate selected scattering processes highlighted in Fig.~3 using identical colors. The QPI data has been symmetrized and a background with a Lorentzian shape has been subtracted. (E-F) show enlarged images of the signature features that appear in the QPI and JDOS for their respective terminations as marked in Fig.3A-D with red triangles. The green circles marks a set of scattering vectors connecting time-reversal pairs of Fermi arcs.  }
\label{fig:Figure4}
\end{figure*}
A clear-cut assignment of the key features in the QPI to scattering between the Fermi arcs is obtained when the relative weight of bulk and surface spectral weight is considered. To do this, we introduce a simplified theoretical surface spectral density, obtained by keeping only the electronic states whose spectral weight at the surface is larger than a certain threshold.  Fig.~4A shows the obtained spectral weight at the Fermi energy for the DH termination and it can be observed that Fermi arcs dominate the resulting spectrum. Fig.~4B presents the corresponding JDOS next to the measured QPI.  Remarkably, the JDOS obtained from the Fermi arcs perfectly captures all the key features experimentally observed and the same applies for the K surface, as shown in Fig.~4C,~D. In comparison, the QPI intensity is more spread out for the DH termination, a characteristic that can be explained based on the different extension of the Fermi arcs. Taking for example the triangular feature, for point-like Fermi arcs on type K surfaces, the QPI is focused at the vertices of the triangle (Fig.~4G,~H) while the additional available scattering vectors for the more elongated Fermi arcs in the DH surface effectively redistribute the QPI intensity towards the interior of the triangle (Fig.~4E,~F). The wave vectors at the vertices of this triangle are all symmetry related:  $\mathbf q^K_2$ and $\mathbf q^K_3$ connect Fermi arcs belonging to different Brillouin zones and are related to each other by reflection symmetry, while $\mathbf q^K_1$ represents the intra-zone scattering and is related to the other two by a lattice translation and reflection. The relative lattice translation symmetry suggests that the former vectors correspond to Umklapp processes, which might in turn explain the experimentally observed asymmetric intensity of these three scattering vectors.

We investigated the energy dispersion of the QPI by studying the QPI in a wide energy range ([-600 meV, 600 meV] and [-50 meV, 50 meV] for type K surfaces), see Figures S3 and S4 for the full data sets.  In Fig.~5A the energy dependence of two features for a K type sample, which proved detectable over a larger energy range than the other details of the QPI, is presented. Here the dispersion of scattering vectors $\mathbf q^K_1$ and $\mathbf q^K_6$ has been extracted from the dataset shown in Fig.~S3 and plotted together with the JDOS calculations (see Fig.~S1). In accordance with the theoretical results, the observed energy dispersion of both wave vectors is linear and $\mathbf q^K_6$ decreases with energy while $\mathbf q^K_1$ does the opposite. A good agreement is also observed at larger positive energies even if the observed QPI features become more diffuse than the features observed closer to the chemical potential. For example, as is shown in Fig.~5B and C for 50~meV and 25~meV, respectively, the spectral weight at  $\mathbf q^K_1$ remains intense and their positions agree well with the JDOS results. Note that data shown in Fig.~5B,~C and Fig.~S4 correspond to a different sample then was previously presented, illustrating that the observed features appear consistently across normal state sample surfaces.

\begin{figure*}
\includegraphics[scale=0.5]{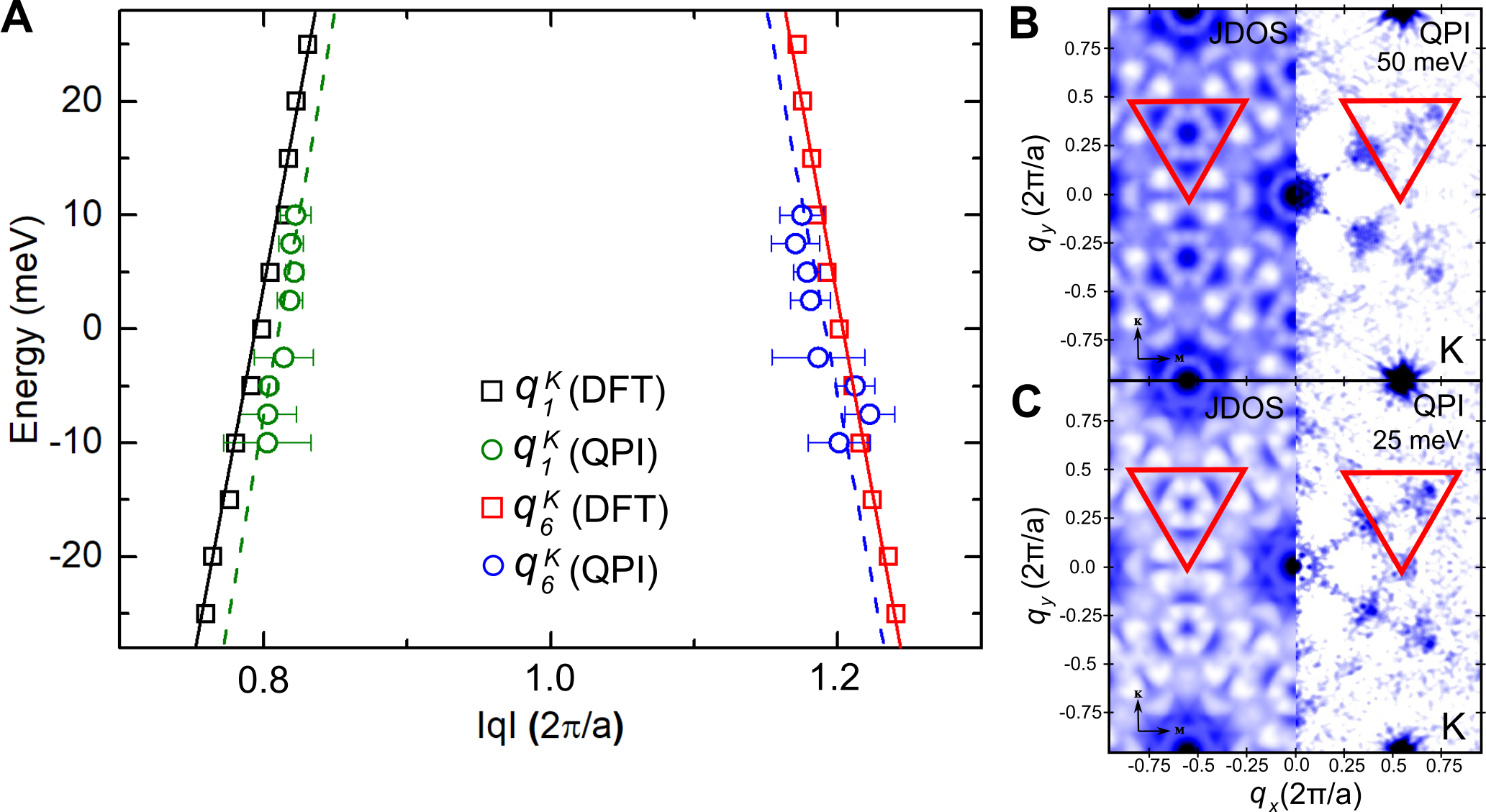}
\caption{(A) Energy dispersion at scattering vectors  $\mathbf q^K_1$ (green circle) and  $\mathbf q^K_6$ (blue circle) recorded for sample 1 at $T=30$ mK. All magnitudes were determined through Lorentz fits of line profiles over corresponding features in Figs. S1 and S3. Error bars present the standard deviation of the fit. Black and red squares mark the corresponding wavevectors extracted in the same way from our DFT calculations.   (B) QPI (right) and JDOS (left)  with 50 meV voltage bias ($I=1.5$ nA, $ U_{\text{rms}}= 2$ meV). (C) Same with a voltage bias of 25 meV ($I=750$ pA, $U_{rms}= 0.3$ meV ). B and C were recorded on sample 3 at $T=5$~K. The red triangle marks the characteristic triangular structure associated with Fermi arcs that was also present at lower energies. The QPI data has been symmetrized and a logarithmic background has been subtracted. }
\label{fig:Figure5}
\end{figure*}

\section{Discussion}

It is remarkable that the QPI features associated with the Fermi arcs are sharper at the Fermi energy than at the Weyl node energy. DFT data suggests that this behavior is connected to particularities of the strong $k_z$ dependence of the electronic structure \cite{Gao2018}. In particular, our calculations indicate that the Fermi arcs are less hybridized with the projection of the bulk bands at the Fermi energy (Fig.~S1 and Fig.~S2). This effective decoupling underlies the experimental manifestation of the Fermi arcs in the QPI pattern.

Our results show that in t-PtBi$_2$ the JDOS can be a useful guide to interpret QPI data, as found previously in Weyl semimetals and other materials \cite{Hoffman2002,Zheng2016,Inoue2016,Kourtis2016,Batabyal2016}. Interestingly, there are as well features in the JDOS with no correspondence in the QPI pattern. This is not unexpected since the phases of Bloch and the matrix elements (such as the spin structure) of the surface Green's function are disregarded in the JDOS approximation. Therefore, non-trivial quantum mechanical interference effects \cite{Mitchell2016} are not reproduced by our calculations. Nonetheless, while the spin texture of the surface Fermi arcs has been shown to hold significant importance to characterize the superconducting state \cite{Vocaturo2024}, we expect to yield meaningful contributions to the QPI only for relatively small $\mathbf k$-vectors and, therefore, the overall agreement with the JDOS is not affected.

Examples of (small) discrepancies are marked by green circles in Fig.~3D and 4D
and can be associated with scattering processes connecting time-reversal Fermi arc pairs, as highlighted by green arrows in Fig.~3F. Thus, our experiments also reflect expected destructive interference imposed by the time-reversal symmetry \cite{Inoue2016,Kourtis2016}.

The layered structure and the strong spin-orbit coupling of all its constituent atoms make t-PtBi$_2$ a Weyl semimetal with rich bulk and surface electronic structures. This richness is rooted in the stacking of decorated honeycomb and kagome-like lattices separated by the Pt plane. This inversion symmetry breaking pattern, in addition to enabling the existence of Weyl nodes, found in ab-initio calculations near the Fermi energy, makes the associated Fermi arcs strongly dependent on the surface termination. This dependence manifests in a different elongation of the Fermi arcs which in turn affects the shape of the associated QPI features.

The striking agreement between the experimental QPI patterns and first-principles calculations
of the topological Fermi arc JDOS (Fig.~4)
provides conclusive evidence of the dominance of the topological Fermi arcs for the surface electronic properties of the Weyl semimetal t-PtBi$_2$.
Our results thus clearly corroborate the notion that the recently reported surface superconductivity \cite{Schimmel2023,Kuibarov2024} is exclusively sustained by the topological Fermi arc states. This underpins the exceptional position of t-PtBi$_2$ among superconducting Weyl semimetals \cite{Bachmann2017,Qi2016,Xing2020,Wang2021} and calls for further experiments to scrutinize its  superconducting phase with respect to topological properties such as Majorana zero modes and potential exploitation for fault-tolerant quantum computation \cite{Kitaev2003,Sato2017}.


%
%
%

\bibliography{PtBi}

\begin{thebibliography}{27}%
\makeatletter
\providecommand \@ifxundefined [1]{%
 \@ifx{#1\undefined}
}%
\providecommand \@ifnum [1]{%
 \ifnum #1\expandafter \@firstoftwo
 \else \expandafter \@secondoftwo
 \fi
}%
\providecommand \@ifx [1]{%
 \ifx #1\expandafter \@firstoftwo
 \else \expandafter \@secondoftwo
 \fi
}%
\providecommand \natexlab [1]{#1}%
\providecommand \enquote  [1]{``#1''}%
\providecommand \bibnamefont  [1]{#1}%
\providecommand \bibfnamefont [1]{#1}%
\providecommand \citenamefont [1]{#1}%
\providecommand \href@noop [0]{\@secondoftwo}%
\providecommand \href [0]{\begingroup \@sanitize@url \@href}%
\providecommand \@href[1]{\@@startlink{#1}\@@href}%
\providecommand \@@href[1]{\endgroup#1\@@endlink}%
\providecommand \@sanitize@url [0]{\catcode `\\12\catcode `\$12\catcode
  `\&12\catcode `\#12\catcode `\^12\catcode `\_12\catcode `\%12\relax}%
\providecommand \@@startlink[1]{}%
\providecommand \@@endlink[0]{}%
\providecommand \url  [0]{\begingroup\@sanitize@url \@url }%
\providecommand \@url [1]{\endgroup\@href {#1}{\urlprefix }}%
\providecommand \urlprefix  [0]{URL }%
\providecommand \Eprint [0]{\href }%
\providecommand \doibase [0]{https://doi.org/}%
\providecommand \selectlanguage [0]{\@gobble}%
\providecommand \bibinfo  [0]{\@secondoftwo}%
\providecommand \bibfield  [0]{\@secondoftwo}%
\providecommand \translation [1]{[#1]}%
\providecommand \BibitemOpen [0]{}%
\providecommand \bibitemStop [0]{}%
\providecommand \bibitemNoStop [0]{.\EOS\space}%
\providecommand \EOS [0]{\spacefactor3000\relax}%
\providecommand \BibitemShut  [1]{\csname bibitem#1\endcsname}%
\let\auto@bib@innerbib\@empty
\bibitem [{\citenamefont {Veyrat}\ \emph {et~al.}(2023)\citenamefont {Veyrat},
  \citenamefont {Labracherie}, \citenamefont {Bashlakov}, \citenamefont
  {Caglieris}, \citenamefont {Facio}, \citenamefont {Shipunov}, \citenamefont
  {Charvin}, \citenamefont {Acharya}, \citenamefont {Naidyuk}, \citenamefont
  {Giraud}, \citenamefont {van~den Brink}, \citenamefont {Büchner},
  \citenamefont {Hess}, \citenamefont {Aswartham},\ and\ \citenamefont
  {Dufouleur}}]{Veyrat2023}%
  \BibitemOpen
  \bibfield  {author} {\bibinfo {author} {\bibfnamefont {A.}~\bibnamefont
  {Veyrat}}, \bibinfo {author} {\bibfnamefont {V.}~\bibnamefont {Labracherie}},
  \bibinfo {author} {\bibfnamefont {D.~L.}\ \bibnamefont {Bashlakov}}, \bibinfo
  {author} {\bibfnamefont {F.}~\bibnamefont {Caglieris}}, \bibinfo {author}
  {\bibfnamefont {J.~I.}\ \bibnamefont {Facio}}, \bibinfo {author}
  {\bibfnamefont {G.}~\bibnamefont {Shipunov}}, \bibinfo {author}
  {\bibfnamefont {T.}~\bibnamefont {Charvin}}, \bibinfo {author} {\bibfnamefont
  {R.}~\bibnamefont {Acharya}}, \bibinfo {author} {\bibfnamefont
  {Y.}~\bibnamefont {Naidyuk}}, \bibinfo {author} {\bibfnamefont
  {R.}~\bibnamefont {Giraud}}, \bibinfo {author} {\bibfnamefont
  {J.}~\bibnamefont {van~den Brink}}, \bibinfo {author} {\bibfnamefont
  {B.}~\bibnamefont {Büchner}}, \bibinfo {author} {\bibfnamefont
  {C.}~\bibnamefont {Hess}}, \bibinfo {author} {\bibfnamefont {S.}~\bibnamefont
  {Aswartham}},\ and\ \bibinfo {author} {\bibfnamefont {J.}~\bibnamefont
  {Dufouleur}},\ }\bibfield  {title} {\bibinfo {title}
  {{Berezinskii-Kosterlitz-Thouless Transition in the Type-I Weyl Semimetal
  PtBi$_2$}},\ }\href@noop {} {\bibfield  {journal} {\bibinfo  {journal} {Nano
  Lett.}\ } (\bibinfo {year} {2023})}\BibitemShut {NoStop}%
\bibitem [{\citenamefont {Shipunov}\ \emph {et~al.}(2020)\citenamefont
  {Shipunov}, \citenamefont {Kovalchuk}, \citenamefont {Piening}, \citenamefont
  {Labracherie}, \citenamefont {Veyrat}, \citenamefont {Wolf}, \citenamefont
  {Lubk}, \citenamefont {Subakti}, \citenamefont {Giraud}, \citenamefont
  {Dufouleur}, \citenamefont {Shokri}, \citenamefont {Caglieris}, \citenamefont
  {Hess}, \citenamefont {Efremov}, \citenamefont {B\"uchner},\ and\
  \citenamefont {Aswartham}}]{Shipunov2020}%
  \BibitemOpen
  \bibfield  {author} {\bibinfo {author} {\bibfnamefont {G.}~\bibnamefont
  {Shipunov}}, \bibinfo {author} {\bibfnamefont {I.}~\bibnamefont {Kovalchuk}},
  \bibinfo {author} {\bibfnamefont {B.~R.}\ \bibnamefont {Piening}}, \bibinfo
  {author} {\bibfnamefont {V.}~\bibnamefont {Labracherie}}, \bibinfo {author}
  {\bibfnamefont {A.}~\bibnamefont {Veyrat}}, \bibinfo {author} {\bibfnamefont
  {D.}~\bibnamefont {Wolf}}, \bibinfo {author} {\bibfnamefont {A.}~\bibnamefont
  {Lubk}}, \bibinfo {author} {\bibfnamefont {S.}~\bibnamefont {Subakti}},
  \bibinfo {author} {\bibfnamefont {R.}~\bibnamefont {Giraud}}, \bibinfo
  {author} {\bibfnamefont {J.}~\bibnamefont {Dufouleur}}, \bibinfo {author}
  {\bibfnamefont {S.}~\bibnamefont {Shokri}}, \bibinfo {author} {\bibfnamefont
  {F.}~\bibnamefont {Caglieris}}, \bibinfo {author} {\bibfnamefont
  {C.}~\bibnamefont {Hess}}, \bibinfo {author} {\bibfnamefont {D.~V.}\
  \bibnamefont {Efremov}}, \bibinfo {author} {\bibfnamefont {B.}~\bibnamefont
  {B\"uchner}},\ and\ \bibinfo {author} {\bibfnamefont {S.}~\bibnamefont
  {Aswartham}},\ }\bibfield  {title} {\bibinfo {title} {{Polymorphic
  ${\mathrm{PtBi}}_{2}$: Growth, structure, and superconducting properties}},\
  }\href@noop {} {\bibfield  {journal} {\bibinfo  {journal} {Phys. Rev.
  Materials}\ }\textbf {\bibinfo {volume} {4}},\ \bibinfo {pages} {124202}
  (\bibinfo {year} {2020})}\BibitemShut {NoStop}%
\bibitem [{\citenamefont {Zabala}\ \emph {et~al.}(2024)\citenamefont {Zabala},
  \citenamefont {Correa}, \citenamefont {Castro},\ and\ \citenamefont
  {Pedrazzini}}]{Zabala2024}%
  \BibitemOpen
  \bibfield  {author} {\bibinfo {author} {\bibfnamefont {J.}~\bibnamefont
  {Zabala}}, \bibinfo {author} {\bibfnamefont {V.~F.}\ \bibnamefont {Correa}},
  \bibinfo {author} {\bibfnamefont {F.~J.}\ \bibnamefont {Castro}},\ and\
  \bibinfo {author} {\bibfnamefont {P.}~\bibnamefont {Pedrazzini}},\ }\bibfield
   {title} {\bibinfo {title} {{Enhanced weak superconductivity in trigonal
  $\gamma$-PtBi$_2$}},\ }\href {https://doi.org/10.1088/1361-648X/ad3878}
  {\bibfield  {journal} {\bibinfo  {journal} {Journal of Physics: Condensed
  Matter}\ }\textbf {\bibinfo {volume} {36}},\ \bibinfo {pages} {285701}
  (\bibinfo {year} {2024})}\BibitemShut {NoStop}%
\bibitem [{\citenamefont {Berezinskii}(1971)}]{Berezinskii1971}%
  \BibitemOpen
  \bibfield  {author} {\bibinfo {author} {\bibfnamefont {V.~L.}\ \bibnamefont
  {Berezinskii}},\ }\bibfield  {title} {\bibinfo {title} {{Destruction of long
  range order in one-dimensional and tw- dimensional Sytems possesing a
  cotinuous symmetry group.}},\ }\href@noop {} {\bibfield  {journal} {\bibinfo
  {journal} {Zh. Eksp. Teor. Fiz.}\ } (\bibinfo {year} {1971})}\BibitemShut
  {NoStop}%
\bibitem [{\citenamefont {Kosterlitz}\ and\ \citenamefont
  {Thouless}(1973)}]{Kosterlitz1973}%
  \BibitemOpen
  \bibfield  {author} {\bibinfo {author} {\bibfnamefont {J.~M.}\ \bibnamefont
  {Kosterlitz}}\ and\ \bibinfo {author} {\bibfnamefont {D.~J.}\ \bibnamefont
  {Thouless}},\ }\bibfield  {title} {\bibinfo {title} {{Ordering, metastability
  and phase transitions in two-dimensional systems}},\ }\href@noop {}
  {\bibfield  {journal} {\bibinfo  {journal} {Journal of Physics C: Solid State
  Physics}\ }\textbf {\bibinfo {volume} {6}},\ \bibinfo {pages} {1181}
  (\bibinfo {year} {1973})}\BibitemShut {NoStop}%
\bibitem [{\citenamefont {Schimmel}\ \emph {et~al.}(2023)\citenamefont
  {Schimmel}, \citenamefont {Fasano}, \citenamefont {Hoffmann}, \citenamefont
  {Puig}, \citenamefont {Shipunov}, \citenamefont {Baumann}, \citenamefont
  {Aswartham}, \citenamefont {Büchner},\ and\ \citenamefont
  {Hess}}]{Schimmel2023}%
  \BibitemOpen
  \bibfield  {author} {\bibinfo {author} {\bibfnamefont {S.}~\bibnamefont
  {Schimmel}}, \bibinfo {author} {\bibfnamefont {Y.}~\bibnamefont {Fasano}},
  \bibinfo {author} {\bibfnamefont {S.}~\bibnamefont {Hoffmann}}, \bibinfo
  {author} {\bibfnamefont {J.}~\bibnamefont {Puig}}, \bibinfo {author}
  {\bibfnamefont {G.}~\bibnamefont {Shipunov}}, \bibinfo {author}
  {\bibfnamefont {D.}~\bibnamefont {Baumann}}, \bibinfo {author} {\bibfnamefont
  {S.}~\bibnamefont {Aswartham}}, \bibinfo {author} {\bibfnamefont
  {B.}~\bibnamefont {Büchner}},\ and\ \bibinfo {author} {\bibfnamefont
  {C.}~\bibnamefont {Hess}},\ }\href
  {https://doi.org/https://doi.org/10.48550/arXiv.2302.08968} {\bibinfo {title}
  {{High-TC surface superconductivity in topological Weyl semimetal
  t-PtBi$_2$}}},\ \bibinfo {howpublished} {arXiv:2302.08968} (\bibinfo {year}
  {2023})\BibitemShut {NoStop}%
\bibitem [{\citenamefont {Kuibarov}\ \emph {et~al.}(2024)\citenamefont
  {Kuibarov}, \citenamefont {Suvorov}, \citenamefont {Vocaturo}, \citenamefont
  {Fedorov}, \citenamefont {Lou}, \citenamefont {Merkwitz}, \citenamefont
  {Voroshnin}, \citenamefont {Facio}, \citenamefont {Koepernik}, \citenamefont
  {Yaresko}, \citenamefont {Shipunov}, \citenamefont {Aswartham}, \citenamefont
  {Brink}, \citenamefont {Büchner},\ and\ \citenamefont
  {Borisenko}}]{Kuibarov2024}%
  \BibitemOpen
  \bibfield  {author} {\bibinfo {author} {\bibfnamefont {A.}~\bibnamefont
  {Kuibarov}}, \bibinfo {author} {\bibfnamefont {O.}~\bibnamefont {Suvorov}},
  \bibinfo {author} {\bibfnamefont {R.}~\bibnamefont {Vocaturo}}, \bibinfo
  {author} {\bibfnamefont {A.}~\bibnamefont {Fedorov}}, \bibinfo {author}
  {\bibfnamefont {R.}~\bibnamefont {Lou}}, \bibinfo {author} {\bibfnamefont
  {L.}~\bibnamefont {Merkwitz}}, \bibinfo {author} {\bibfnamefont
  {V.}~\bibnamefont {Voroshnin}}, \bibinfo {author} {\bibfnamefont {J.~I.}\
  \bibnamefont {Facio}}, \bibinfo {author} {\bibfnamefont {K.}~\bibnamefont
  {Koepernik}}, \bibinfo {author} {\bibfnamefont {A.}~\bibnamefont {Yaresko}},
  \bibinfo {author} {\bibfnamefont {G.}~\bibnamefont {Shipunov}}, \bibinfo
  {author} {\bibfnamefont {S.}~\bibnamefont {Aswartham}}, \bibinfo {author}
  {\bibfnamefont {J.~v.~d.}\ \bibnamefont {Brink}}, \bibinfo {author}
  {\bibfnamefont {B.}~\bibnamefont {Büchner}},\ and\ \bibinfo {author}
  {\bibfnamefont {S.}~\bibnamefont {Borisenko}},\ }\bibfield  {title} {\bibinfo
  {title} {{Evidence of superconducting Fermi arcs}},\ }\href
  {https://doi.org/10.1038/s41586-023-06977-7} {\bibfield  {journal} {\bibinfo
  {journal} {Nature}\ }\textbf {\bibinfo {volume} {626}},\ \bibinfo {pages}
  {294} (\bibinfo {year} {2024})}\BibitemShut {NoStop}%
\bibitem [{\citenamefont {Bachmann}\ \emph {et~al.}(2017)\citenamefont
  {Bachmann}, \citenamefont {Nair}, \citenamefont {Flicker}, \citenamefont
  {Ilan}, \citenamefont {Meng}, \citenamefont {Ghimire}, \citenamefont {Bauer},
  \citenamefont {Ronning}, \citenamefont {Analytis},\ and\ \citenamefont
  {Moll}}]{Bachmann2017}%
  \BibitemOpen
  \bibfield  {author} {\bibinfo {author} {\bibfnamefont {M.~D.}\ \bibnamefont
  {Bachmann}}, \bibinfo {author} {\bibfnamefont {N.}~\bibnamefont {Nair}},
  \bibinfo {author} {\bibfnamefont {F.}~\bibnamefont {Flicker}}, \bibinfo
  {author} {\bibfnamefont {R.}~\bibnamefont {Ilan}}, \bibinfo {author}
  {\bibfnamefont {T.}~\bibnamefont {Meng}}, \bibinfo {author} {\bibfnamefont
  {N.~J.}\ \bibnamefont {Ghimire}}, \bibinfo {author} {\bibfnamefont {E.~D.}\
  \bibnamefont {Bauer}}, \bibinfo {author} {\bibfnamefont {F.}~\bibnamefont
  {Ronning}}, \bibinfo {author} {\bibfnamefont {J.~G.}\ \bibnamefont
  {Analytis}},\ and\ \bibinfo {author} {\bibfnamefont {P.~J.~W.}\ \bibnamefont
  {Moll}},\ }\bibfield  {title} {\bibinfo {title} {{Inducing superconductivity
  in Weyl semimetal microstructures by selective ion sputtering}},\ }\href@noop
  {} {\bibfield  {journal} {\bibinfo  {journal} {Science Advances}\ }\textbf
  {\bibinfo {volume} {3}},\ \bibinfo {pages} {e1602983} (\bibinfo {year}
  {2017})}\BibitemShut {NoStop}%
\bibitem [{\citenamefont {Qi}\ \emph {et~al.}(2016)\citenamefont {Qi},
  \citenamefont {Naumov}, \citenamefont {Ali}, \citenamefont {Rajamathi},
  \citenamefont {Schnelle}, \citenamefont {Barkalov}, \citenamefont {Hanfland},
  \citenamefont {Wu}, \citenamefont {Shekhar}, \citenamefont {Sun},
  \citenamefont {Süß}, \citenamefont {Schmidt}, \citenamefont {Schwarz},
  \citenamefont {Pippel}, \citenamefont {Werner}, \citenamefont {Hillebrand},
  \citenamefont {Förster}, \citenamefont {Kampert}, \citenamefont {Parkin},
  \citenamefont {Cava}, \citenamefont {Felser}, \citenamefont {Yan},\ and\
  \citenamefont {Medvedev}}]{Qi2016}%
  \BibitemOpen
  \bibfield  {author} {\bibinfo {author} {\bibfnamefont {Y.}~\bibnamefont
  {Qi}}, \bibinfo {author} {\bibfnamefont {P.~G.}\ \bibnamefont {Naumov}},
  \bibinfo {author} {\bibfnamefont {M.~N.}\ \bibnamefont {Ali}}, \bibinfo
  {author} {\bibfnamefont {C.~R.}\ \bibnamefont {Rajamathi}}, \bibinfo {author}
  {\bibfnamefont {W.}~\bibnamefont {Schnelle}}, \bibinfo {author}
  {\bibfnamefont {O.}~\bibnamefont {Barkalov}}, \bibinfo {author}
  {\bibfnamefont {M.}~\bibnamefont {Hanfland}}, \bibinfo {author}
  {\bibfnamefont {S.-C.}\ \bibnamefont {Wu}}, \bibinfo {author} {\bibfnamefont
  {C.}~\bibnamefont {Shekhar}}, \bibinfo {author} {\bibfnamefont
  {Y.}~\bibnamefont {Sun}}, \bibinfo {author} {\bibfnamefont {V.}~\bibnamefont
  {Süß}}, \bibinfo {author} {\bibfnamefont {M.}~\bibnamefont {Schmidt}},
  \bibinfo {author} {\bibfnamefont {U.}~\bibnamefont {Schwarz}}, \bibinfo
  {author} {\bibfnamefont {E.}~\bibnamefont {Pippel}}, \bibinfo {author}
  {\bibfnamefont {P.}~\bibnamefont {Werner}}, \bibinfo {author} {\bibfnamefont
  {R.}~\bibnamefont {Hillebrand}}, \bibinfo {author} {\bibfnamefont
  {T.}~\bibnamefont {Förster}}, \bibinfo {author} {\bibfnamefont
  {E.}~\bibnamefont {Kampert}}, \bibinfo {author} {\bibfnamefont
  {S.}~\bibnamefont {Parkin}}, \bibinfo {author} {\bibfnamefont {R.~J.}\
  \bibnamefont {Cava}}, \bibinfo {author} {\bibfnamefont {C.}~\bibnamefont
  {Felser}}, \bibinfo {author} {\bibfnamefont {B.}~\bibnamefont {Yan}},\ and\
  \bibinfo {author} {\bibfnamefont {S.~A.}\ \bibnamefont {Medvedev}},\
  }\bibfield  {title} {\bibinfo {title} {{Superconductivity in Weyl semimetal
  candidate MoTe$_2$}},\ }\href {https://doi.org/10.1038/ncomms11038}
  {\bibfield  {journal} {\bibinfo  {journal} {Nature Communications}\ }\textbf
  {\bibinfo {volume} {7}},\ \bibinfo {pages} {11038} (\bibinfo {year}
  {2016})}\BibitemShut {NoStop}%
\bibitem [{\citenamefont {Xing}\ \emph {et~al.}(2020)\citenamefont {Xing},
  \citenamefont {Shao}, \citenamefont {Ge}, \citenamefont {Luo}, \citenamefont
  {Wang}, \citenamefont {Zhu}, \citenamefont {Liu}, \citenamefont {Wang},
  \citenamefont {Zhao}, \citenamefont {Yan}, \citenamefont {Mandrus},
  \citenamefont {Yan}, \citenamefont {Liu}, \citenamefont {Pan},\ and\
  \citenamefont {Wang}}]{Xing2020}%
  \BibitemOpen
  \bibfield  {author} {\bibinfo {author} {\bibfnamefont {Y.}~\bibnamefont
  {Xing}}, \bibinfo {author} {\bibfnamefont {Z.}~\bibnamefont {Shao}}, \bibinfo
  {author} {\bibfnamefont {J.}~\bibnamefont {Ge}}, \bibinfo {author}
  {\bibfnamefont {J.}~\bibnamefont {Luo}}, \bibinfo {author} {\bibfnamefont
  {J.}~\bibnamefont {Wang}}, \bibinfo {author} {\bibfnamefont {Z.}~\bibnamefont
  {Zhu}}, \bibinfo {author} {\bibfnamefont {J.}~\bibnamefont {Liu}}, \bibinfo
  {author} {\bibfnamefont {Y.}~\bibnamefont {Wang}}, \bibinfo {author}
  {\bibfnamefont {Z.}~\bibnamefont {Zhao}}, \bibinfo {author} {\bibfnamefont
  {J.}~\bibnamefont {Yan}}, \bibinfo {author} {\bibfnamefont {D.}~\bibnamefont
  {Mandrus}}, \bibinfo {author} {\bibfnamefont {B.}~\bibnamefont {Yan}},
  \bibinfo {author} {\bibfnamefont {X.-J.}\ \bibnamefont {Liu}}, \bibinfo
  {author} {\bibfnamefont {M.}~\bibnamefont {Pan}},\ and\ \bibinfo {author}
  {\bibfnamefont {J.}~\bibnamefont {Wang}},\ }\bibfield  {title} {\bibinfo
  {title} {{Surface superconductivity in the type II Weyl semimetal
  TaIrTe$_4$}},\ }\href {https://doi.org/10.1093/nsr/nwz204} {\bibfield
  {journal} {\bibinfo  {journal} {National Science Review}\ }\textbf {\bibinfo
  {volume} {7}},\ \bibinfo {pages} {579} (\bibinfo {year} {2020})}\BibitemShut
  {NoStop}%
\bibitem [{\citenamefont {Wang}\ \emph {et~al.}(2021)\citenamefont {Wang},
  \citenamefont {Olivares}, \citenamefont {Namiki}, \citenamefont {Pareek},
  \citenamefont {Dani}, \citenamefont {Sasagawa}, \citenamefont {Madhavan},\
  and\ \citenamefont {Okada}}]{Wang2021}%
  \BibitemOpen
  \bibfield  {author} {\bibinfo {author} {\bibfnamefont {Z.}~\bibnamefont
  {Wang}}, \bibinfo {author} {\bibfnamefont {J.}~\bibnamefont {Olivares}},
  \bibinfo {author} {\bibfnamefont {H.}~\bibnamefont {Namiki}}, \bibinfo
  {author} {\bibfnamefont {V.}~\bibnamefont {Pareek}}, \bibinfo {author}
  {\bibfnamefont {K.}~\bibnamefont {Dani}}, \bibinfo {author} {\bibfnamefont
  {T.}~\bibnamefont {Sasagawa}}, \bibinfo {author} {\bibfnamefont
  {V.}~\bibnamefont {Madhavan}},\ and\ \bibinfo {author} {\bibfnamefont
  {Y.}~\bibnamefont {Okada}},\ }\bibfield  {title} {\bibinfo {title}
  {{Visualizing superconductivity in a doped Weyl semimetal with broken
  inversion symmetry}},\ }\href {https://doi.org/10.1103/PhysRevB.104.115102}
  {\bibfield  {journal} {\bibinfo  {journal} {Phys. Rev. B}\ }\textbf {\bibinfo
  {volume} {104}},\ \bibinfo {pages} {115102} (\bibinfo {year}
  {2021})}\BibitemShut {NoStop}%
\bibitem [{\citenamefont {Thirupathaiah}\ \emph {et~al.}(2018)\citenamefont
  {Thirupathaiah}, \citenamefont {Kushnirenko}, \citenamefont {Haubold},
  \citenamefont {Fedorov}, \citenamefont {Rienks}, \citenamefont {Kim},
  \citenamefont {Yaresko}, \citenamefont {Blum}, \citenamefont {Aswartham},
  \citenamefont {B\"uchner},\ and\ \citenamefont
  {Borisenko}}]{Thirupathaiah2018}%
  \BibitemOpen
  \bibfield  {author} {\bibinfo {author} {\bibfnamefont {S.}~\bibnamefont
  {Thirupathaiah}}, \bibinfo {author} {\bibfnamefont {Y.}~\bibnamefont
  {Kushnirenko}}, \bibinfo {author} {\bibfnamefont {E.}~\bibnamefont
  {Haubold}}, \bibinfo {author} {\bibfnamefont {A.~V.}\ \bibnamefont
  {Fedorov}}, \bibinfo {author} {\bibfnamefont {E.~D.~L.}\ \bibnamefont
  {Rienks}}, \bibinfo {author} {\bibfnamefont {T.~K.}\ \bibnamefont {Kim}},
  \bibinfo {author} {\bibfnamefont {A.~N.}\ \bibnamefont {Yaresko}}, \bibinfo
  {author} {\bibfnamefont {C.~G.~F.}\ \bibnamefont {Blum}}, \bibinfo {author}
  {\bibfnamefont {S.}~\bibnamefont {Aswartham}}, \bibinfo {author}
  {\bibfnamefont {B.}~\bibnamefont {B\"uchner}},\ and\ \bibinfo {author}
  {\bibfnamefont {S.~V.}\ \bibnamefont {Borisenko}},\ }\bibfield  {title}
  {\bibinfo {title} {{Possible origin of linear magnetoresistance: Observation
  of Dirac surface states in layered ${\mathrm{PtBi}}_{2}$}},\ }\href@noop {}
  {\bibfield  {journal} {\bibinfo  {journal} {Phys. Rev. B}\ }\textbf {\bibinfo
  {volume} {97}},\ \bibinfo {pages} {035133} (\bibinfo {year}
  {2018})}\BibitemShut {NoStop}%
\bibitem [{\citenamefont {Gao}\ \emph {et~al.}(2018)\citenamefont {Gao},
  \citenamefont {Zhu}, \citenamefont {Zheng}, \citenamefont {Wu}, \citenamefont
  {Zhang}, \citenamefont {Xi}, \citenamefont {Zhang}, \citenamefont {Zhang},
  \citenamefont {Hao}, \citenamefont {Ning},\ and\ \citenamefont
  {Tian}}]{Gao2018}%
  \BibitemOpen
  \bibfield  {author} {\bibinfo {author} {\bibfnamefont {W.}~\bibnamefont
  {Gao}}, \bibinfo {author} {\bibfnamefont {X.}~\bibnamefont {Zhu}}, \bibinfo
  {author} {\bibfnamefont {F.}~\bibnamefont {Zheng}}, \bibinfo {author}
  {\bibfnamefont {M.}~\bibnamefont {Wu}}, \bibinfo {author} {\bibfnamefont
  {J.}~\bibnamefont {Zhang}}, \bibinfo {author} {\bibfnamefont
  {C.}~\bibnamefont {Xi}}, \bibinfo {author} {\bibfnamefont {P.}~\bibnamefont
  {Zhang}}, \bibinfo {author} {\bibfnamefont {Y.}~\bibnamefont {Zhang}},
  \bibinfo {author} {\bibfnamefont {N.}~\bibnamefont {Hao}}, \bibinfo {author}
  {\bibfnamefont {W.}~\bibnamefont {Ning}},\ and\ \bibinfo {author}
  {\bibfnamefont {M.}~\bibnamefont {Tian}},\ }\bibfield  {title} {\bibinfo
  {title} {{A possible candidate for triply degenerate point fermions in
  trigonal layered PtBi$_2$}},\ }\href@noop {} {\bibfield  {journal} {\bibinfo
  {journal} {Nature Communications}\ }\textbf {\bibinfo {volume} {9}},\
  \bibinfo {pages} {3249} (\bibinfo {year} {2018})}\BibitemShut {NoStop}%
\bibitem [{\citenamefont {Hoffman}\ \emph {et~al.}(2002)\citenamefont
  {Hoffman}, \citenamefont {McElroy}, \citenamefont {Lee}, \citenamefont
  {Lang}, \citenamefont {Eisaki}, \citenamefont {Uchida},\ and\ \citenamefont
  {Davis}}]{Hoffman2002}%
  \BibitemOpen
  \bibfield  {author} {\bibinfo {author} {\bibfnamefont {J.~E.}\ \bibnamefont
  {Hoffman}}, \bibinfo {author} {\bibfnamefont {K.}~\bibnamefont {McElroy}},
  \bibinfo {author} {\bibfnamefont {D.-H.}\ \bibnamefont {Lee}}, \bibinfo
  {author} {\bibfnamefont {K.~M.}\ \bibnamefont {Lang}}, \bibinfo {author}
  {\bibfnamefont {H.}~\bibnamefont {Eisaki}}, \bibinfo {author} {\bibfnamefont
  {S.}~\bibnamefont {Uchida}},\ and\ \bibinfo {author} {\bibfnamefont {J.~C.}\
  \bibnamefont {Davis}},\ }\bibfield  {title} {\bibinfo {title} {{Imaging
  Quasiparticle Interference in Bi$_2$Sr$_2$CaCu$_2$O$_{8+\delta}$}},\
  }\href@noop {} {\bibfield  {journal} {\bibinfo  {journal} {Science}\ }\textbf
  {\bibinfo {volume} {297}},\ \bibinfo {pages} {1148} (\bibinfo {year}
  {2002})}\BibitemShut {NoStop}%
\bibitem [{\citenamefont {Zheng}\ and\ \citenamefont
  {Zahid~Hasan}(2018)}]{Zheng2018}%
  \BibitemOpen
  \bibfield  {author} {\bibinfo {author} {\bibfnamefont {H.}~\bibnamefont
  {Zheng}}\ and\ \bibinfo {author} {\bibfnamefont {M.}~\bibnamefont
  {Zahid~Hasan}},\ }\bibfield  {title} {\bibinfo {title} {{Quasiparticle
  interference on type-I and type-II Weyl semimetal surfaces: a review}},\
  }\href@noop {} {\bibfield  {journal} {\bibinfo  {journal} {Advances in
  Physics: X}\ }\textbf {\bibinfo {volume} {3}},\ \bibinfo {pages} {1466661}
  (\bibinfo {year} {2018})}\BibitemShut {NoStop}%
\bibitem [{\citenamefont {Nie}\ \emph {et~al.}(2020)\citenamefont {Nie},
  \citenamefont {Li}, \citenamefont {Yang}, \citenamefont {Zhu}, \citenamefont
  {Xu}, \citenamefont {Yang}, \citenamefont {Zheng}, \citenamefont {Guan},
  \citenamefont {Wang}, \citenamefont {Li}, \citenamefont {Liu}, \citenamefont
  {Li}, \citenamefont {Zhang}, \citenamefont {Shi}, \citenamefont {Zheng},\
  and\ \citenamefont {Jia}}]{Nie2020}%
  \BibitemOpen
  \bibfield  {author} {\bibinfo {author} {\bibfnamefont {X.-A.}\ \bibnamefont
  {Nie}}, \bibinfo {author} {\bibfnamefont {S.}~\bibnamefont {Li}}, \bibinfo
  {author} {\bibfnamefont {M.}~\bibnamefont {Yang}}, \bibinfo {author}
  {\bibfnamefont {Z.}~\bibnamefont {Zhu}}, \bibinfo {author} {\bibfnamefont
  {H.-K.}\ \bibnamefont {Xu}}, \bibinfo {author} {\bibfnamefont
  {X.}~\bibnamefont {Yang}}, \bibinfo {author} {\bibfnamefont {F.}~\bibnamefont
  {Zheng}}, \bibinfo {author} {\bibfnamefont {D.}~\bibnamefont {Guan}},
  \bibinfo {author} {\bibfnamefont {S.}~\bibnamefont {Wang}}, \bibinfo {author}
  {\bibfnamefont {Y.-Y.}\ \bibnamefont {Li}}, \bibinfo {author} {\bibfnamefont
  {C.}~\bibnamefont {Liu}}, \bibinfo {author} {\bibfnamefont {J.}~\bibnamefont
  {Li}}, \bibinfo {author} {\bibfnamefont {P.}~\bibnamefont {Zhang}}, \bibinfo
  {author} {\bibfnamefont {Y.}~\bibnamefont {Shi}}, \bibinfo {author}
  {\bibfnamefont {H.}~\bibnamefont {Zheng}},\ and\ \bibinfo {author}
  {\bibfnamefont {J.}~\bibnamefont {Jia}},\ }\bibfield  {title} {\bibinfo
  {title} {{Robust Hot Electron and Multiple Topological Insulator States in
  PtBi$_2$}},\ }\href@noop {} {\bibfield  {journal} {\bibinfo  {journal} {ACS
  Nano}\ }\textbf {\bibinfo {volume} {14}},\ \bibinfo {pages} {2366} (\bibinfo
  {year} {2020})}\BibitemShut {NoStop}%
\bibitem [{\citenamefont {Zheng}\ \emph {et~al.}(2016)\citenamefont {Zheng},
  \citenamefont {Bian}, \citenamefont {Chang}, \citenamefont {Lu},
  \citenamefont {Xu}, \citenamefont {Wang}, \citenamefont {Chang},
  \citenamefont {Zhang}, \citenamefont {Belopolski}, \citenamefont {Alidoust},
  \citenamefont {Sanchez}, \citenamefont {Song}, \citenamefont {Jeng},
  \citenamefont {Yao}, \citenamefont {Bansil}, \citenamefont {Jia},
  \citenamefont {Lin},\ and\ \citenamefont {Hasan}}]{Zheng2016}%
  \BibitemOpen
  \bibfield  {author} {\bibinfo {author} {\bibfnamefont {H.}~\bibnamefont
  {Zheng}}, \bibinfo {author} {\bibfnamefont {G.}~\bibnamefont {Bian}},
  \bibinfo {author} {\bibfnamefont {G.}~\bibnamefont {Chang}}, \bibinfo
  {author} {\bibfnamefont {H.}~\bibnamefont {Lu}}, \bibinfo {author}
  {\bibfnamefont {S.-Y.}\ \bibnamefont {Xu}}, \bibinfo {author} {\bibfnamefont
  {G.}~\bibnamefont {Wang}}, \bibinfo {author} {\bibfnamefont {T.-R.}\
  \bibnamefont {Chang}}, \bibinfo {author} {\bibfnamefont {S.}~\bibnamefont
  {Zhang}}, \bibinfo {author} {\bibfnamefont {I.}~\bibnamefont {Belopolski}},
  \bibinfo {author} {\bibfnamefont {N.}~\bibnamefont {Alidoust}}, \bibinfo
  {author} {\bibfnamefont {D.~S.}\ \bibnamefont {Sanchez}}, \bibinfo {author}
  {\bibfnamefont {F.}~\bibnamefont {Song}}, \bibinfo {author} {\bibfnamefont
  {H.-T.}\ \bibnamefont {Jeng}}, \bibinfo {author} {\bibfnamefont
  {N.}~\bibnamefont {Yao}}, \bibinfo {author} {\bibfnamefont {A.}~\bibnamefont
  {Bansil}}, \bibinfo {author} {\bibfnamefont {S.}~\bibnamefont {Jia}},
  \bibinfo {author} {\bibfnamefont {H.}~\bibnamefont {Lin}},\ and\ \bibinfo
  {author} {\bibfnamefont {M.~Z.}\ \bibnamefont {Hasan}},\ }\bibfield  {title}
  {\bibinfo {title} {{Atomic-Scale Visualization of Quasiparticle Interference
  on a Type-II Weyl Semimetal Surface}},\ }\href
  {https://doi.org/10.1103/PhysRevLett.117.266804} {\bibfield  {journal}
  {\bibinfo  {journal} {Phys. Rev. Lett.}\ }\textbf {\bibinfo {volume} {117}},\
  \bibinfo {pages} {266804} (\bibinfo {year} {2016})}\BibitemShut {NoStop}%
\bibitem [{\citenamefont {Inoue}\ \emph {et~al.}(2016)\citenamefont {Inoue},
  \citenamefont {Gyenis}, \citenamefont {Wang}, \citenamefont {Li},
  \citenamefont {Oh}, \citenamefont {Jiang}, \citenamefont {Ni}, \citenamefont
  {Bernevig},\ and\ \citenamefont {Yazdani}}]{Inoue2016}%
  \BibitemOpen
  \bibfield  {author} {\bibinfo {author} {\bibfnamefont {H.}~\bibnamefont
  {Inoue}}, \bibinfo {author} {\bibfnamefont {A.}~\bibnamefont {Gyenis}},
  \bibinfo {author} {\bibfnamefont {Z.}~\bibnamefont {Wang}}, \bibinfo {author}
  {\bibfnamefont {J.}~\bibnamefont {Li}}, \bibinfo {author} {\bibfnamefont
  {S.~W.}\ \bibnamefont {Oh}}, \bibinfo {author} {\bibfnamefont
  {S.}~\bibnamefont {Jiang}}, \bibinfo {author} {\bibfnamefont
  {N.}~\bibnamefont {Ni}}, \bibinfo {author} {\bibfnamefont {B.~A.}\
  \bibnamefont {Bernevig}},\ and\ \bibinfo {author} {\bibfnamefont
  {A.}~\bibnamefont {Yazdani}},\ }\bibfield  {title} {\bibinfo {title}
  {{Quasiparticle interference of the Fermi arcs and surface-bulk connectivity
  of a Weyl semimetal}},\ }\href@noop {} {\bibfield  {journal} {\bibinfo
  {journal} {Science}\ }\textbf {\bibinfo {volume} {351}},\ \bibinfo {pages}
  {1184} (\bibinfo {year} {2016})}\BibitemShut {NoStop}%
\bibitem [{\citenamefont {Kourtis}\ \emph {et~al.}(2016)\citenamefont
  {Kourtis}, \citenamefont {Li}, \citenamefont {Wang}, \citenamefont
  {Yazdani},\ and\ \citenamefont {Bernevig}}]{Kourtis2016}%
  \BibitemOpen
  \bibfield  {author} {\bibinfo {author} {\bibfnamefont {S.}~\bibnamefont
  {Kourtis}}, \bibinfo {author} {\bibfnamefont {J.}~\bibnamefont {Li}},
  \bibinfo {author} {\bibfnamefont {Z.}~\bibnamefont {Wang}}, \bibinfo {author}
  {\bibfnamefont {A.}~\bibnamefont {Yazdani}},\ and\ \bibinfo {author}
  {\bibfnamefont {B.~A.}\ \bibnamefont {Bernevig}},\ }\bibfield  {title}
  {\bibinfo {title} {{Universal signatures of Fermi arcs in quasiparticle
  interference on the surface of Weyl semimetals}},\ }\href@noop {} {\bibfield
  {journal} {\bibinfo  {journal} {Phys. Rev. B}\ }\textbf {\bibinfo {volume}
  {93}} (\bibinfo {year} {2016})}\BibitemShut {NoStop}%
\bibitem [{\citenamefont {Batabyal}\ \emph {et~al.}(2016)\citenamefont
  {Batabyal}, \citenamefont {Morali}, \citenamefont {Avraham}, \citenamefont
  {Sun}, \citenamefont {Schmidt}, \citenamefont {Felser}, \citenamefont
  {A.Stern}, \citenamefont {Yan},\ and\ \citenamefont
  {Beidenkopf}}]{Batabyal2016}%
  \BibitemOpen
  \bibfield  {author} {\bibinfo {author} {\bibfnamefont {R.}~\bibnamefont
  {Batabyal}}, \bibinfo {author} {\bibfnamefont {N.}~\bibnamefont {Morali}},
  \bibinfo {author} {\bibfnamefont {N.}~\bibnamefont {Avraham}}, \bibinfo
  {author} {\bibfnamefont {Y.}~\bibnamefont {Sun}}, \bibinfo {author}
  {\bibfnamefont {M.}~\bibnamefont {Schmidt}}, \bibinfo {author} {\bibfnamefont
  {C.}~\bibnamefont {Felser}}, \bibinfo {author} {\bibnamefont {A.Stern}},
  \bibinfo {author} {\bibfnamefont {B.}~\bibnamefont {Yan}},\ and\ \bibinfo
  {author} {\bibfnamefont {H.}~\bibnamefont {Beidenkopf}},\ }\bibfield  {title}
  {\bibinfo {title} {{Visualizing weakly bound surface Fermi arcs and their
  correspondence to bulk Weyl fermions}},\ }\href@noop {} {\bibfield  {journal}
  {\bibinfo  {journal} {Science Advances}\ }\textbf {\bibinfo {volume} {2}}
  (\bibinfo {year} {2016})}\BibitemShut {NoStop}%
\bibitem [{\citenamefont {Mitchell}\ and\ \citenamefont
  {Fritz}(2016)}]{Mitchell2016}%
  \BibitemOpen
  \bibfield  {author} {\bibinfo {author} {\bibfnamefont {A.~K.}\ \bibnamefont
  {Mitchell}}\ and\ \bibinfo {author} {\bibfnamefont {L.}~\bibnamefont
  {Fritz}},\ }\bibfield  {title} {\bibinfo {title} {{Signatures of Weyl
  semimetals in quasiparticle interference}},\ }\href@noop {} {\bibfield
  {journal} {\bibinfo  {journal} {Phys. Rev. B}\ }\textbf {\bibinfo {volume}
  {93}},\ \bibinfo {pages} {035137} (\bibinfo {year} {2016})}\BibitemShut
  {NoStop}%
\bibitem [{\citenamefont {Vocaturo}\ \emph {et~al.}(2024)\citenamefont
  {Vocaturo}, \citenamefont {Koepernik}, \citenamefont {Facio}, \citenamefont
  {Timm}, \citenamefont {Fulga}, \citenamefont {Janson},\ and\ \citenamefont
  {van~den Brink}}]{Vocaturo2024}%
  \BibitemOpen
  \bibfield  {author} {\bibinfo {author} {\bibfnamefont {R.}~\bibnamefont
  {Vocaturo}}, \bibinfo {author} {\bibfnamefont {K.}~\bibnamefont {Koepernik}},
  \bibinfo {author} {\bibfnamefont {J.~I.}\ \bibnamefont {Facio}}, \bibinfo
  {author} {\bibfnamefont {C.}~\bibnamefont {Timm}}, \bibinfo {author}
  {\bibfnamefont {I.~C.}\ \bibnamefont {Fulga}}, \bibinfo {author}
  {\bibfnamefont {O.}~\bibnamefont {Janson}},\ and\ \bibinfo {author}
  {\bibfnamefont {J.}~\bibnamefont {van~den Brink}},\ }\href
  {https://doi.org/https://doi.org/10.48550/arXiv.2404.19606} {\bibinfo {title}
  {{Electronic structure of the surface superconducting Weyl semimetal
  PtBi$_2$}}},\ \bibinfo {howpublished} {arXiv:2404.19606v1} (\bibinfo {year}
  {2024})\BibitemShut {NoStop}%
\bibitem [{\citenamefont {Kitaev}(2003)}]{Kitaev2003}%
  \BibitemOpen
  \bibfield  {author} {\bibinfo {author} {\bibfnamefont {A.~Y.}\ \bibnamefont
  {Kitaev}},\ }\bibfield  {title} {\bibinfo {title} {{Fault-tolerant quantum
  computation by anyons}},\ }\href@noop {} {\bibfield  {journal} {\bibinfo
  {journal} {Annals of Physics}\ }\textbf {\bibinfo {volume} {303}},\ \bibinfo
  {pages} {2} (\bibinfo {year} {2003})}\BibitemShut {NoStop}%
\bibitem [{\citenamefont {Sato}\ and\ \citenamefont {Ando}(2017)}]{Sato2017}%
  \BibitemOpen
  \bibfield  {author} {\bibinfo {author} {\bibfnamefont {M.}~\bibnamefont
  {Sato}}\ and\ \bibinfo {author} {\bibfnamefont {Y.}~\bibnamefont {Ando}},\
  }\bibfield  {title} {\bibinfo {title} {{Topological superconductors: a
  review}},\ }\href@noop {} {\bibfield  {journal} {\bibinfo  {journal} {Reports
  on Progress in Physics}\ }\textbf {\bibinfo {volume} {80}},\ \bibinfo {pages}
  {076501} (\bibinfo {year} {2017})}\BibitemShut {NoStop}%
\bibitem [{\citenamefont {Schlegel}\ \emph {et~al.}(2014)\citenamefont
  {Schlegel}, \citenamefont {H\"anke}, \citenamefont {Baumann}, \citenamefont
  {Kaiser}, \citenamefont {Nag}, \citenamefont {Voigtl\"ander}, \citenamefont
  {Lindackers}, \citenamefont {B\"uchner},\ and\ \citenamefont
  {Hess}}]{Schlegel2014}%
  \BibitemOpen
  \bibfield  {author} {\bibinfo {author} {\bibfnamefont {R.}~\bibnamefont
  {Schlegel}}, \bibinfo {author} {\bibfnamefont {T.}~\bibnamefont {H\"anke}},
  \bibinfo {author} {\bibfnamefont {D.}~\bibnamefont {Baumann}}, \bibinfo
  {author} {\bibfnamefont {M.}~\bibnamefont {Kaiser}}, \bibinfo {author}
  {\bibfnamefont {P.~K.}\ \bibnamefont {Nag}}, \bibinfo {author} {\bibfnamefont
  {R.}~\bibnamefont {Voigtl\"ander}}, \bibinfo {author} {\bibfnamefont
  {D.}~\bibnamefont {Lindackers}}, \bibinfo {author} {\bibfnamefont
  {B.}~\bibnamefont {B\"uchner}},\ and\ \bibinfo {author} {\bibfnamefont
  {C.}~\bibnamefont {Hess}},\ }\bibfield  {title} {\bibinfo {title} {{Design
  and properties of a cryogenic dip-stick scanning tunneling microscope with
  capacitive coarse approach control}},\ }\href@noop {} {\bibfield  {journal}
  {\bibinfo  {journal} {Review of Scientific Instruments}\ }\textbf {\bibinfo
  {volume} {85}},\ \bibinfo {pages} {013706} (\bibinfo {year}
  {2014})}\BibitemShut {NoStop}%
\bibitem [{\citenamefont {Koepernik}\ and\ \citenamefont
  {Eschrig}(1999)}]{Koepernik1999}%
  \BibitemOpen
  \bibfield  {author} {\bibinfo {author} {\bibfnamefont {K.}~\bibnamefont
  {Koepernik}}\ and\ \bibinfo {author} {\bibfnamefont {H.}~\bibnamefont
  {Eschrig}},\ }\bibfield  {title} {\bibinfo {title} {{Full-potential
  nonorthogonal local-orbital minimum-basis band-structure scheme}},\ }\href
  {https://doi.org/10.1103/PhysRevB.59.1743} {\bibfield  {journal} {\bibinfo
  {journal} {Phys. Rev. B}\ }\textbf {\bibinfo {volume} {59}},\ \bibinfo
  {pages} {1743} (\bibinfo {year} {1999})}\BibitemShut {NoStop}%
\bibitem [{\citenamefont {Perdew}\ \emph {et~al.}(1996)\citenamefont {Perdew},
  \citenamefont {Burke},\ and\ \citenamefont {Ernzerhof}}]{Perdew1996}%
  \BibitemOpen
  \bibfield  {author} {\bibinfo {author} {\bibfnamefont {J.~P.}\ \bibnamefont
  {Perdew}}, \bibinfo {author} {\bibfnamefont {K.}~\bibnamefont {Burke}},\ and\
  \bibinfo {author} {\bibfnamefont {M.}~\bibnamefont {Ernzerhof}},\ }\bibfield
  {title} {\bibinfo {title} {{Generalized Gradient Approximation Made
  Simple}},\ }\href {https://doi.org/10.1103/PhysRevLett.77.3865} {\bibfield
  {journal} {\bibinfo  {journal} {Phys. Rev. Lett.}\ }\textbf {\bibinfo
  {volume} {77}},\ \bibinfo {pages} {3865} (\bibinfo {year}
  {1996})}\BibitemShut {NoStop}%
\end{thebibliography}%

\section{Methods}

\subsection{Sample preparation and STM investigations}
Stoichiometric crystals of t-PtBi$_2$ were grown via the self-flux method. A detailed description of the crystal growth as well as thorough characterization of the samples crystal structure and stoichiometry is provided in \cite{Shipunov2020}. The samples were mounted inside the STM and cleaved under ultra-high vacuum conditions ensuring atomically clean surfaces. In total 3 different samples have been investigated using two home built STM setups. Sample 1 of the type K termination was analyzed with an ultra-low-temperature STM \cite{Schimmel2023} capable of maintaining a base temperature of 30 mK over long periods of time by means of dilution-refrigeration. Another device based on a dip-stick design \cite{Schlegel2014} was used for samples 2 and 3, providing stable measuring conditions at a temperature of 5 K. Here sample 2 is of type DH termination while sample 3 represents another example for a type K surface. All three samples exhibited a normal sate during all measurements. For all spectroscopic maps presented here lock-in amplification was used to intrinsically measure the differential conductance. All measurements were performed in constant current mode with the feedback loop active. Tip stabilization parameters defined by current set point $I$ and bias voltage $U$ as well as the modulation voltage $U_{\text{rms}}$, applied by the lock-in amplifier to the bias voltage at a frequency of $f = 1.111$ kHz are given in the caption of all figures in the main text.

\subsection{Data visualization}
QPI pattern are derived trough fast Fourier transformation (FFT) of the acquired d\textit{I}/d\textit{U} maps. All QPI patterns have been symmetrized following the FFT to increase contrast. Additionally, from the QPI data shown in Figs.~3 and 5 a Lorentz shaped (Fig.~3) or logarithmic (Fig.~5) background has been subtracted depending on magnitude and the consequently better coverage of the central peak in the FFT-images.

\subsection{Density functional theory}
Fully relativistic DFT calculations were performed using the FPLO package with the standard basis setting \cite{Koepernik1999} and the Perdew-Burke-Ernzerhof version of the generalized gradient approximation \cite{Perdew1996}. Brillouin zone integrations were done with tetrahedron method on a k-mesh having $12\times12\times13$ subdivisions. The electronic structure at surfaces was studied using a tight-binding Hamiltonian in the Wannier basis that comprised Pt-$5d$ and $6s$ as well Bi-$6p$ states. For parametrization, we constructed symmetry-conserving maximally projected Wannier functions as implemented in FPLO. The calculations are based on the experimentally determined crystal structure \cite{Shipunov2020}. For completeness, we repeated our calculations for the centrosymmetric structure of t-PtBi$_2$ considered in \cite{Thirupathaiah2018} where Weyl nodes are absent as it possesses both time-reversal and inversion symmetries. The resulting JDOS is markedly different, see Fig.~S5.\\

\section{Acknowledgments}

We thank Ulrike Nietzsche for technical assistance. This work received support from the Deutsche Forschungsgemeinschaft (grants 500507880, AS 523/4–1 and SFB 1143) as well as the Dresden-Würzburg Cluster of Excellence (EXC 2147). Furthermore, this project received funding from the European Research Council (grant 647276 – MARS – ERC-2014-CoG), Alexander von Humboldt Stiftung,  ANPCyT (grants PICT 2017/2182,  PICT 2018/01509 and PICT 2019/00371) and Conicet (grant PIP 2021/1848).

\section{Author contributions}
C.H, B.B., J.V.D.B. designed and supervised the research, S.S, D.B., and C.H. developed the experimental methodology, S.H., S.S., J.P., Y.F. performed STM measurements, R.V., J.I.F., O.J., and J.V.D.B performed first-principle and JDOS calculations, G.S. and S.A. grew samples, all authors discussed the data analysis and interpretation; S.H., S.S., Y.F., J.I.F., and C.H. wrote the paper.

\section{Competing interests}
The authors declare no competing interests.\\

\noindent\textbf{Correspondence and requests for materials} should be addressed to Christian Hess.

\end{document}